# Student difficulties with the probability distribution for measuring energy and position in quantum mechanics

Chandralekha Singh and Emily Marshman
*Department of Physics and Astronomy, University of Pittsburgh, Pittsburgh, PA 15260*

**Abstract.** We have been investigating the difficulties that students in upper-level undergraduate courses have in determining the probability distribution for measuring energy and position as a function of time when the initial wave function is explicitly given. We administered written free-response and multiple-choice questions to investigate these difficulties. We find that students have common difficulties with these concepts. We discuss some of the findings.

**PACS:** 1.40Fk

## I. INTRODUCTION

Learning quantum mechanics is challenging [1-6]. Measurement is what connects abstract quantum theory to experiments. According to the quantum postulates that students learn, measurement outcomes are probabilistic. Issues related to measurement of energy and position for a given quantum system are central to the quantum formalism. Here, we discuss difficulties that students in upper-level undergraduate courses have in computing the probability distribution for measuring energy and position as a function of time when the initial wave function is explicitly given.

## II. METHODOLOGY AND RESULTS

Student difficulties with the probability distribution for measuring energy and position were investigated by administering free-response and multiple-choice surveys to upper-level undergraduate students and conducting individual interviews with a subset of students. Here, we discuss some questions in the position representation and others in Dirac notation involving measurement probabilities which were administered to upper-level students after traditional instruction at six universities in the U.S. Four multiple-choice questions administered to 210 students explicitly asked about energy and position measurement probabilities at time $t = 0$ and at a later time $t$ when the initial wave function $\Psi(x, t = 0)$ was provided. Three questions about the measurement probabilities were posed using Dirac notation to 272 students. The individual interviews employed a think-aloud protocol to better understand the rationale for student responses. During the semi-structured interviews, students were asked to "think aloud" while answering questions. They first read the questions and reasoned about them without interruptions except that they were prompted to think aloud if they were quiet for a long time. After students had finished answering a particular question to the best of their ability, we asked them to further clarify and elaborate issues that they had not clearly addressed on their own earlier.

### A. Difficulties with measurement probabilities for a given wave function $\Psi(x, t = 0)$

In four multiple-choice questions as follows, students ($N = 210$) were given the initial wave function at $t = 0$ and were asked about the probability distribution for measuring energy and position for a particle in a one-dimensional infinite square well between $x = 0$ and $x = a$ at time $t = 0$ and time $t > 0$. Students were told that the stationary state wave functions for this system are $\psi_n(x) = \sqrt{2/a} \sin(n\pi x/a)$ and the allowed energies are $E_n = n^2\pi^2\hbar^2/2ma^2$ where $= 1,2,3 \dots$ . (Correct answers are bolded).

*Question 1.* The wave function at time $t = 0$ is $\Psi(x, 0) = Ax(a − x)$ for $0 \leq x \leq a$, where $A$ is a suitable normalization constant. Choose all of the following statements that are correct <u>at time $t = 0$</u>:
(1) If you measure the position of the particle at time <u>$t = 0$</u>, the probability density for measuring $x$ is $|Ax(a − x)|^2$.
(2) If you measure the energy of the system at time <u>$t = 0$</u>, the probability of obtaining $E_1$ is $\left|\int_0^a \psi_1^*(x)Ax(a − x)\,dx\right|^2$.
(3) If you measure the position of the particle at time <u>$t = 0$</u>, the probability of obtaining a value between $x$ and $x + dx$ is $\int_x^{x+dx} x|\Psi(x,0)|^2\,dx$.
A. 1 only  B. 3 only  **C. 1 and 2 only**  D. 1 and 3 only
E. All of the above

*Question 2.* The wave function at time $t = 0$ is $\Psi(x, 0) = Ax(a − x)$ for $0 \leq x \leq a$, where $A$ is a suitable normalization constant. Choose all of the following statements that are correct <u>at time $t > 0$</u>:

(1) If you measure the position of the particle <u>after a time $t$</u>, the probability density for measuring $x$ is $|Ax(a − x)|^2$.
(2) If you measure the energy of the system <u>after a time $t$</u>, the probability of obtaining $E_1$ is $\left|\int_0^a \psi_1^*(x)Ax(a − x)dx\right|^2$.

(3) If you measure the position of the particle <u>after a time t</u>, the probability of obtaining a value between $x$ and $x + dx$ is $\int_x^{x+dx} x|\Psi(x,0)|^2 \, dx$.

A. None of the above  B. 1 only  **C. 2 only**  D. 3 only
E. 1 and 3 only

Questions 3 and 4 below are identical to questions 1 and 2, respectively, except for the choice of the initial wave function at time $t = 0$.

***Question 3.*** The wave function at time $t = 0$ is $\Psi(x,0) = (\psi_1(x) + \psi_2(x))/\sqrt{2}$. Choose all of the following statements that are correct <u>at time $t = 0$</u>:
(1) If you measure the position of the particle at time <u>$t = 0$</u>, the probability density for measuring $x$ is $\left|(\psi_1(x) + \psi_2(x))/\sqrt{2}\right|^2$.
(2) If you measure the energy of the system at time <u>$t = 0$</u>, the probability of obtaining $E_1$ is $\left|\int_0^a \psi_1^*(x)(\psi_1(x) + \psi_2(x))/\sqrt{2} \, dx\right|^2$.
(3) If you measure the position of the particle at time <u>$t = 0$</u>, the probability of obtaining a value between $x$ and $x + dx$ is $\int_x^{x+dx} x|\Psi(x,0)|^2 \, dx$.

A. 1 only  B. 3 only  **C. 1 and 2 only**  D. 1 and 3 only
E. All of the above

***Question 4.*** The wave function at time $t = 0$ is $\Psi(x,0) = (\psi_1(x) + \psi_2(x))/\sqrt{2}$. Choose all of the following statements that are correct <u>at time $t > 0$</u>:
(1) If you measure the position of the particle <u>after a time t</u>, the probability density for measuring $x$ is $\left|(\psi_1(x) + \psi_2(x))/\sqrt{2}\right|^2$.
(2) If you measure the energy of the system <u>after a time t</u>, the probability of obtaining $E_1$ is $\left|\int_0^a \psi_1^*(x)\big((\psi_1(x) + \psi_2(x))/\sqrt{2}\big)dx\right|^2$.
(3) If you measure the position of the particle <u>after a time t</u>, the probability of obtaining a value between $x$ and $x + dx$ is $\int_x^{x+dx} x|\Psi(x,0)|^2 \, dx$.

A. None of the above  B. 1 only  **C. 2 only**  D. 3 only
E. 1 and 3 only

We will also compare student performance on question 2 with the following question posed about the probability density which does not explicitly mention "position measurement" in the problem statement:

***Question 5.*** Consider the following wave function at time $t = 0$: $\Psi(x,0) = Ax(a-x)$ for $0 \le x \le a$, where $A$ is a suitable normalization constant. Which one of the following is the probability density $|\Psi(x,t)|^2$, at time $t > 0$?

A. $|\Psi(x,t)|^2 = |A|^2 x^2 (a-x)^2 \cos^2\left(\frac{Et}{\hbar}\right)$, where $E$ is the expectation value of energy.
B. $|\Psi(x,t)|^2 = |A|^2 x^2 (a-x)^2 e^{\frac{-2iEt}{\hbar}}$, where $E$ is the expectation value of energy.
C. $|\Psi(x,t)|^2 = |A|^2 x^2 (a-x)^2 \sin^2\left(\frac{Et}{\hbar}\right)$, where $E$ is the expectation value of energy.
D. $|\Psi(x,t)|^2 = |A|^2 x^2 (a-x)^2$, which is time-independent.
E. *None of the above.*

Below, we summarize the common difficulties:

**Confusing expectation value of position with probability distribution for measuring position at time $t = 0$:** According to Born's interpretation, the probability of measuring the particle's position between $x$ and $x + dx$ at time $t = 0$ is $|\Psi(x,0)|^2 dx$. On questions involving the probability density for measuring $x$ at time $t = 0$ (questions 1 and 3), approximately 80% of the students correctly recognized that the probability density for measuring $x$ is $|Ax(a-x)|^2$ (or $\left|(\psi_1(x) + \psi_2(x))/\sqrt{2}\right|^2$). However, Table I shows that approximately 60% of the students incorrectly responded that the probability of measuring position is $\int_x^{x+dx} x|\Psi(x,0)|^2 dx$. This dichotomy indicates that many students do not discern the connection between probability density and the probability of measuring position between $x$ and $x + dx$, i.e., one can multiply the probability density $|\Psi(x,0)|^2$ by infinitesimal interval $dx$ to obtain the probability of measuring the position in a narrow range between $x$ and $x + dx$. Interviews suggest that some students who thought that statement (3) in questions 1 and 3 was correct confused the probability of measuring position with the expectation value of position (although the integral in statement (3) is not from $x = 0$ to $x = a$, necessary for the expectation value).

**Incorrect assumption that the probability density for measuring $x$ is time-independent:** Table I shows that on questions involving the probability density for measuring $x$ at time $t > 0$ (questions 2 and 4), 28% of the students agreed with statement (1). This type of response indicates that students have difficulty reasoning about how the wave function will evolve in time according to the Hamiltonian $\hat{H}$ of the system in a non-trivial manner and the probability of measuring an observable such as position whose corresponding operator does not commute with $\hat{H}$ will depend on time.

**Context-dependent responses involving probability density at time $t > 0$:** Although there is no explicit mention of a position measurement in question 5, students were explicitly asked to select the probability density $|\Psi(x,t)|^2$ at time $t > 0$ for the same initial wave function as in questions 1 and 2. Thus, while questions 2 and 5 are posed differently, an expert would recognize that answer choice (D) in question 5 and statement (1) in question 2 are conceptually very similar. Table I shows that on question 5, 48% of the students incorrectly selected the answer choice (D) $|\Psi(x,t)|^2 = |A|^2 x^2 (a-x)^2$, which is time-independent. This indicates that students incorrectly assumed that the probability density is found by taking the absolute square of the initial wave function, even if the quantum state is not initially an energy eigenstate. Table I shows that the percentage of students (48%) who incorrectly assumed that the probability density for measuring $x$ depends on time on question 5 is significantly higher than the percentage of students (28%) who made the same incorrect assumption on question 2 and selected statement (1). Interviews suggest that since the expression for the probability density was explicitly stated in question 5, i.e., $|\Psi(x,t)|^2$, some students used it as a cue to find the answer by simply squaring the initial wave function, i.e., $|A|^2 x^2 (a-x)^2$. On question 2 (and even on question 4), students were less likely to choose expressions such as $|Ax(a-x)|^2$ (statement 1) as correct for the probability density at time $t > 0$ because they were not cued with the expression for the probability density, $|\Psi(x,t)|^2$. Moreover, since question 5 did not explicitly mention position measurement, some interviewed students did not realize that the probability density $|\Psi(x,t)|^2$ in question 5 is the probability density *for measuring position* in statement (1) in question 2. These types of discrepancies demonstrate how student responses are sensitive to the context and how the questions are posed. An expert in quantum mechanics would not be distracted by the fact that the expression for probability density, $|\Psi(x,t)|^2$, was included in the problem statement for question 5. However, students who are developing expertise in quantum mechanics may respond differently to questions which are worded slightly differently since they have not developed a coherent knowledge structure [7]. Their knowledge structure is locally consistent and certain cues may prime them to answer incorrectly.

**Difficulties with the probability distribution for an energy measurement at time $t = 0$:** The probability for measuring energy $E_1$ given an initial wave function $\Psi(x,0)$ is $\left|\int_0^a \psi_1^*(x)\Psi(x,0)dx\right|^2$. Table I shows that on question 1, approximately 50% of students did not recognize that the probability of measuring energy $E_1$ is $\left|\int_0^a \psi_1^*(x)Ax(a-x)\,dx\right|^2$. On question 3, the initial wave function is $\Psi(x,0) = (\psi_1(x) + \psi_2(x))/\sqrt{2}$. An expert would immediately recognize that the probability of measuring $E_1$ is ½ and can be obtained using $\left|\int_0^a \psi_1^*(x)((\psi_1(x)+\psi_2(x))/\sqrt{2})dx\right|^2$ as in statement (2) in question 3. However, Table I shows that half of the students did not recognize that statement (2) in question 3 is correct. Interviews suggest that even students who recognize that the probability of measuring energy $E_1$ is ½ for this wave function, which is an equal superposition of ground and first excited states, did not recognize that the integral in statement (2) in question 3 gives the component of the quantum state along the ground state and is related to the energy measurement amplitude.

**Incorrect assumption that the probability distribution for an energy measurement depends on time:** Questions 2 and 4 also investigated student understanding of probability of measuring energy at time $t > 0$ (statement (2)). Table I shows that 77% of the students did not select statement (2) in questions 2 and 4 as true indicating that they have difficulty with the fact that energy is conserved for a quantum system for which the Hamiltonian does not depend on time. In other words, the probability of obtaining energy $E_1$ does not depend on time because energy is a constant of motion. Interviews suggest that students have difficulty with why the probability density for measuring position depends on time but the probability of measuring a particular value of energy is time-independent for these systems.

**Table I.** Percentages of students displaying difficulties with position and energy measurements for a given $\Psi(x,0)$.

| Difficulty | % |
|---|---|
| Confusing "expectation value" with the probability of measuring position at time $t = 0$, i.e., claiming that the probability of obtaining a value between $x$ and $x + dx$ is $\int_x^{x+dx} x|\Psi(x,0)|^2\,dx$ (questions 1 and 3) | 60 |
| Incorrect assumption that probability density for measuring $x$ is time independent (questions 2 and 4) | 28 |
| Incorrect assumption that the probability density, $|\Psi(x,t)|^2$, is time independent (question 5) | 48 |
| Difficulty with the probability distribution for an energy measurement at time $t = 0$ (questions 1 and 3) | 50 |
| Incorrect assumption that the probability distribution for energy measurement depends on time (questions 2 and 4) | 77 |

**Giving consistently incorrect responses to analogous questions for measurements made at time $t = 0$:** Questions 1 and 3 are analogous because the initial wave function is not an eigenstate of the Hamiltonian operator and the measurements of position and energy are made at time $t = 0$. Table II shows that 72% of the students answered questions 1 and 3 consistently (e.g., if a student selected answer choice "C" in question 1, he/she also selected answer choice "C" in question 3). This consistency indicates that they recognize the analogous nature of questions 1 and 3. However, only 17% of them answered questions 1 and 3 both consistently and correctly (see Table II). Of those who answered questions 1 and 3 consistently but incorrectly, 45% incorrectly claimed that the probability of measuring position is $\int_x^{x+dx} x|\Psi(x,0)|^2 dx$ and 22% did not recognize that the probability of measuring $E_1$ is $\left|\int_0^a \psi_1^*(x)Ax(a-x)\,dx\right|^2$ or $\left|\int_0^a \psi_1^*(x)((\psi_1(x)+\psi_2(x))/\sqrt{2})dx\right|^2$ in question 1 and 3, respectively.

**Table II.** Percentages of students who consistently answered questions about position and energy measurements

| | |
|---|---|
| Consistent answers to Questions 1 and 3 ($t = 0$) | 72 |
| Correct consistent answers to Questions 1 and 3 ($t = 0$) | 17 |
| Consistent answers to Questions 2 and 4 ($t > 0$) | 69 |
| Correct consistent answers to Questions 2 and 4 ($t > 0$) | 10 |

**Giving consistently incorrect responses to analogous questions for measurements made at time $t > 0$:** Questions 2 and 4 are analogous because the initial wave function is not an eigenstate of the Hamiltonian operator and the measurements of position and energy are made at time

$t > 0$. Table II shows that 69% of the students provided consistent answers on questions 2 and 4. However, only 10% answered questions 2 and 4 consistently and correctly. The most common incorrect but consistent answer was (A), none of the above. This choice indicates that these students correctly recognize that the probability distribution for measuring position depends on time but do not realize that probability distribution for measuring energy does not depend on time since energy is a constant of motion.

### C. Performance on measurement probabilities for questions involving Dirac notation

Dirac notation is an elegant representation which can simplify analysis. Below, we discuss findings from questions involving measurement probabilities in Dirac notation.

**Proficiency in recognizing the probability distribution for a position measurement in Dirac notation:** Students were asked to evaluate the correctness of the following statement: D1. *If you measure the position of the particle in the state $|\Psi\rangle$, the probability of finding the particle between $x$ and $x + dx$ is $|\langle x|\Psi\rangle|^2 dx$*. Table III shows that 85% correctly recognized that the statement is true.

**Proficiency in recognizing the probability distribution for an energy measurement in Dirac notation:** Students were told that $|n\rangle$ is an energy eigenstate corresponding to eigenvalue $E_n$ and asked to evaluate the correctness of the statement: D2. *If you measure the energy of the system in the state $|\Psi\rangle$, the probability of obtaining $E_n$ and collapsing the state to $|n\rangle$ is $|\langle n|\Psi\rangle|^2$*. Table III shows that 91% of the students correctly recognized that the statement is true.

**Dirac notation may be useful to guide student learning:** Students were told that an operator $\hat{Q}$ corresponding to a physical observable $Q$ has a continuous non-degenerate spectrum of eigenvalues and the states $\{|q\rangle\}$ are eigenstates of $\hat{Q}$ with eigenvalues $q$. They were asked to evaluate the correctness of the following two (correct) statements in the same question: D3 (I) "*If you measure $Q$ at time $t = 0$, the probability of obtaining an outcome between $q$ and $q + dq$ is $|\langle q|\Psi\rangle|^2 dq$*" and D3 (II) "*If you measure $Q$ at time $t = 0$, the probability of obtaining an outcome between $q$ and $q + dq$ is $\left|\int_{-\infty}^{\infty} e_q^*(x)\Psi(x)dx\right|^2 dq$ in which $e_q(x)$ and $\Psi(x)$ are the wave functions in position representation corresponding to states $|q\rangle$ and $|\Psi\rangle$, respectively.*" Table III shows that 67% and 60% of the students, respectively, correctly recognized that both statements are true. The statements D3 (I) and D3 (II) have the same underlying concepts about measurement probabilities except that statement D3 (I) is in Dirac notation and statement D3 (II) is in position representation. Student performance on questions 1-4 (~20% correct on each question) about probabilities of energy and position measurement in position representation discussed earlier are worse than on statement D3 (II) here even though it is about the measurement probability of a generic observable $Q$. Interviews suggest that students sometimes took advantage of statement D3 (I) in Dirac notation to determine that statement D3 (II) in position representation was correct. The fact that comparable large percentages of students in written surveys also recognize that both statements D3 (I) and D3 (II) are correct (i.e., 67% and 60%) further indicates that students may use statements posed in Dirac notation as a scaffold to determine probability distribution for measurements in position representation.

**Table III.** Percentages of students who correctly answered questions D1-D3 about probability distribution of measurements.

| Question | % |
|---|---|
| D1. *If you measure the position of the particle in the state $|\Psi\rangle$, the probability of finding the particle between $x$ and $x + dx$ is $|\langle x|\Psi\rangle|^2 dx$* | 85 |
| D2. *If you measure the energy of the system in the state $|\Psi\rangle$, the probability of obtaining $E_n$ and collapsing the state to $|n\rangle$ is $|\langle n|\Psi\rangle|^2$* | 91 |
| D3 (I). *If you measure $Q$ at time $t = 0$, the probability of obtaining an outcome between $q$ and $q + dq$ is $|\langle q|\Psi\rangle|^2 dq$* | 67 |
| D3 (II). *If you measure $Q$ at time $t = 0$, the probability of obtaining an outcome between $q$ and $q + dq$ is $\left|\int_{-\infty}^{\infty} e_q^*(x)\Psi(x)dx\right|^2 dq$ in which $e_q(x)$ and $\Psi(x)$ are the wave functions in position representation corresponding to states $|q\rangle$ and $|\Psi\rangle$, respectively* | 60 |

### III. SUMMARY

Students exhibit many common difficulties with the probability distribution for measuring position and energy as a function of time which often stem from the difficulty in discriminating between related concepts. We find that sometimes students took advantage of questions using Dirac notation to answer questions about the probability distribution without Dirac notation. If students become proficient in using Dirac notation, they may even become more adept at determining probabilities of measuring observables in position representation using scaffolding in which questions in Dirac notation precede those in the position representation.

### ACKNOWLEDGEMENTS

We thank the National Science Foundation for award PHY-1202909.